\documentclass[aps,prd,11pt,notitlepage,superscriptaddress,nofootinbib,tightenlines]{revtex4-1}
\usepackage{graphicx}
\usepackage{amssymb} %maths
\usepackage{amsmath} %maths
\usepackage[utf8]{inputenc} %useful to type directly diacritic characters
\usepackage{slashed}
\usepackage{bbm} %double stroke letters
\usepackage{turnstile}

\newcommand{\Tr}{\mbox{Tr}}

\newcommand{\dt}{d\tau}

\begin{document}

\title{An efficient method to compute the residual phase on a Lefschetz thimble}

\author{M.~Cristoforetti}
\affiliation{ECT$^\star$/FBK. Strada delle tabarelle, 286 -- I-38123 Trento, Italy.}
\affiliation{LISC/FBK. Via Sommarive, 18 -- I-38123 Trento, Italy.}
\author{F.~Di Renzo}
\author{G.~Eruzzi}
\affiliation{Universit\`{a} di Parma and INFN gruppo collegato di Parma. \\ Viale G.P. Usberti, 7/A  --
I-43124 Parma, Italy.}
\author{A.~Mukherjee}
\affiliation{ECT$^\star$/FBK. Strada delle tabarelle, 286 -- I-38123 Trento, Italy.}
\affiliation{LISC/FBK. Via Sommarive, 18 -- I-38123 Trento, Italy.}
\author{C.~Schmidt}
\affiliation{Universit\"at Bielefeld, Fakult\"at f\"ur Physik, Postfach 100131 -- D-33501 Bielefeld, Germany.}
\author{L.~Scorzato}
\affiliation{INFN-TIFPA. Via Sommarive, 14 -- I-38123 Trento, Italy.}
\author{C.~Torrero}
\affiliation{Aix-Marseillle Universit{\'e}, CNRS, CPT, UMR 7332 -- F-13288 Marseille, France.}

\begin{abstract}
We propose an efficient method to compute the so-called {\em residual phase} that appears when performing Monte
Carlo calculations on a Lefschetz thimble.  The method is stochastic and its cost scales linearly with the physical
volume, linearly with the number of stochastic estimators and quadratically with the length of the extra dimension
along the gradient flow.  This is a drastic improvement over previous estimates of the cost of computing the
residual phase.  We also report on basic tests of correctness and scaling of the code.
\end{abstract}

\maketitle

\section{Introduction}

A Lefschetz thimble \cite{Pham1983,Witten:2010cx} has been recently proposed as a tool to regularize quantum field
theories (QFTs) and statistical systems (at least near criticality), in order to evade the infamous {\em sign
  problem} \cite{Cristoforetti:2012su,Cristoforetti:2013wha,Mukherjee:2013aga,Fujii:2013sra,Cristoforetti:2013qaa}.
In the original proposal \cite{Cristoforetti:2012su}, one major difficulty of the approach was envisaged in the
calculation of the so-called {\em residual phase}, that appears in the measure term when the thimble is not a flat
manifold.  This is a potential problem both because it threatens, in principle, to reintroduce a sign problem, and
because its computation was expected to be very expensive (scaling like $n^3$, where $n$ is the number of degrees
of freedom of the original system).

Actually, there are good reasons to expect that the residual phase does not reintroduce a sign problem, although
they cannot be considered conclusive.  First, the residual phase is completely neglected when one computes the
asymptotic expansion around the saddle point that defines the thimble, which is expected to be a reasonable
approximation in many cases.  Second, the thimble does not oscillate unpredictably.  Instead, its orientation
smoothly interpolates between the directions of steepest descent at the saddle point (which are determined by the
quadratic part of the action) and the asymptotic directions of convergence (which are determined by the highest
degree of the interaction).  In general, one can achieve very strange behaviors, by tuning the parameters of an
action, but this is not expected to be the generic case.  Third, the residual phase tends to deviate substantially
from its value at the saddle point only on configurations that are correspondingly suppressed.  Hence, we can
expect that the thimble realizes a strong correlation between {\em phase} and {\em weight}, which is exactly what
is missing on typical cases of difficult sign problem.

Although the previous arguments are merely qualitative, the best (and also quantitative) evidence that the residual
phase does not reintroduce a sign problem is provided by the very precise computation performed in
\cite{Fujii:2013sra} for a complex scalar theory with $\phi^4$ interactions in 4 dimensions.  In fact, it was shown
that the real part of the average residual phase is systematically larger than $0.99$, for all the parameter values
studied there\footnote{It is worthwhile noting that the residual phase introduces sizable corrections when one
  considers the same action, but in zero dimensions \cite{Eruzzi2013,Aarts:2013fpa}.  Although a precise comparison
  is impossible, this suggests a suppression of the contribution of configurations with large residual phases in
  presence of many degrees of freedom.}.  None of these considerations allows us to conclude that the residual
phase will not introduce a sign problem in other regimes or other models.  However, they are certainly sufficient
to motivate further investigation of this approach, including the search for more efficient strategies to compute
the residual phase.

In this paper, we first review how the residual phase appears in the Lefschetz thimble approach, and then we
propose a new method to compute it numerically, with much better scaling properties than the naive method proposed
in \cite{Cristoforetti:2012su}.  The method exploits the properties of the thimble and standard numerical
techniques.  In particular, we compute a trace over space-time indices with stochastic estimators, which ensure
acceptable computational costs and easy parallelization.  Finally, we present some basic tests of the new method on
small lattices.

\section{Definition of the Residual Phase}

In order to understand how the residual phase appears in an integration on a Lefschetz thimble, consider the
integral:
\begin{equation}
\label{eq:int-orig}
\int_{\mathbb{R}^n} f(x) \, \prod_{i=1}^n dx_i.
\end{equation}
The Lefschetz thimble approach leads us to complexify $f(x)$ into $f(z)$, with $z\in\mathbb{C}^n$, and substitute
formula (\ref{eq:int-orig}) with the integral (see Eq.~(1) of \cite{Pham1983}):
\begin{equation}
\label{eq:int}
\int_{\Gamma} f(z) \, dz_1 \wedge \ldots \wedge dz_n,
\end{equation}
where $\Gamma$ is a Lefschetz thimble, and $d^n z:=dz_1 \wedge \ldots \wedge dz_n$ is a form of precisely the right
degree to integrate a manifold of real dimension $n$ in $\mathbb{C}^n$, as it is indeed the dimension of the
manifold $\Gamma$.  (Note that $d^n z$ is not the standard volume form in $\mathbb{C}^n$, which is, instead, $dz_1
\wedge d\bar{z}_1 \wedge \ldots \wedge dz_n \wedge d\bar{z}_n$.)

In a generic point $\zeta\in\Gamma$, the form $d^n z$ and the tangent space $T_{\zeta} \Gamma$ are not parallel.
In order to evaluate the integral (\ref{eq:int}), we need to express it as an ordinary integral in $\mathbb{R}^n$.
To this end, we must change the coordinates from the canonical basis of $\mathbb{C}^n$ (dual to the forms $dz_i$,
$i=1\ldots n$) into a basis of $T_{\zeta} \Gamma$ (let us call such basis $u^{(1)}, \ldots, u^{(n)}$)\footnote{We
  will see, in the next section, that this can be accomplished through a unitary transformation}.  Let $U$ be the
$n\times n$ complex matrix whose columns are the vectors of the basis $u^{(i)}$.

The change of basis can be realized, locally, with a chart $\varphi:N\subset\Gamma \rightarrow \mathbb{R}^n$,
defined on a neighborhood $N\subset\Gamma$ of $\zeta$.  For instance, we can define $\varphi$ as:
\begin{equation}
\varphi(\zeta+\sum_i u^{(i)} y_i) = y + O(y^2) \in \mathbb{R}^n.
\end{equation}
Then the integral (\ref{eq:int}) becomes:
\begin{equation}
\label{eq:int1}
\int_{N} f(z) \, dz_1 \wedge \ldots \wedge dz_n = 
\int_{\varphi(N)} f(\varphi^{-1}(y)) \, \det U(\varphi^{-1}(y)) \, \prod_i dy_i.
\end{equation}

The integral (\ref{eq:int1}) can be performed by Monte Carlo methods.  For this, we need to sample the points in
$\Gamma$ uniformly according to the measure induced by the standard hermitian metric of $\mathbb{C}^n$
(equivalently, the Euclidean metric in $\mathbb{R}^{2n}$), while taking into account the determinant of $U$.  In
the algorithm of \cite{Cristoforetti:2012su}, the metric enters only in the computation of the length of the random
noise vectors, where, indeed, the Euclidean metric in $\mathbb{R}^{2n}$ is used.  This ensures that this algorithm
samples $\Gamma$ uniformly according to the correct measure.  Therefore, we are left with the computation of $\det
U$, which is the topic of the rest of this paper\footnote{Note that in \cite{Mukherjee:2013aga} the residual
  determinant is not exactly the same as the one defined above.  The case of \cite{Mukherjee:2013aga} is discussed
  in the Appendix.}.

\section{Tangent space at the saddle point}
\label{sec:saddle}

It is important to observe that there is a special matrix $J\in M(\mathbb{R}^{2n})$ (almost complex structure) that
represents, in $\mathbb{R}^{2n}$, the multiplication by $i$ in $\mathbb{C}^{n}$. Its form is:
\begin{equation}
J=\left(
\begin{array}{cc}
0 & 1_n \\
-1_n & 0\\
\end{array} \right).
\end{equation}
The matrix $J$ anti-commutes with the Hessian\footnote{We use $i,j$ for multi-indices that include also the
  real/imaginary part of $z$, for all $n$ degrees of freedom.  Hence, $H(z)$ is a $2n\times 2n$ real symmetric
  matrix} $H(z)=\partial_{i,j}^2 S_R(z)$ for each $z$.  This implies that $J$ transforms any eigenvector of $H(z)$
with eigenvalue $\lambda$ into another eigenvector with eigenvalue $-\lambda$.

The thimble is well defined only if the Hessian is non-degenerate at the saddle point $\zeta_0$, and we assume that
this is the case in the following.  Let $V_+$ be the $2n\times n$ real matrix whose columns are the eigenvectors of
$H(\zeta_0)$ with positive eigenvalues and define
\begin{equation}
\label{eq:V-}
V_-:=JV_+.
\end{equation}  
We can define a matrix $U_+$ by the $n$ complex column vectors: $u_h^{(i)}:=v_{R,h}^{(i)} + i v_{I,h}^{(i)}$,
$i,h=1,\ldots,n$.  In matrix notation we can write:
\begin{equation}
\label{eq:P}
U_+=PV_+, \qquad P=(1_n,i \, 1_n).
\end{equation}
Now $U_+$ is unitary.  In fact,
\begin{eqnarray*}
\sum_h \bar{u}^{(i)}_h u^{(j)}_h &=& \sum_h (u^{(i)}_{R,h} - i u^{(i)}_{I,h}) (u^{(j)}_{R,h} + i u^{(j)}_{I,h}) 
= 
\sum_h (u^{(i)}_{R,h})^2 + (u^{(i)}_{I,h})^2 + i (u^{(i)}_{R,h} u^{(j)}_{I,h}  - u^{(i)}_{I,h} u^{(j)}_{R,h}) 
=\\
&=&
(v^{(i)})^2  + i (v^{(i)} J v^{(j)}). \\
\end{eqnarray*}
The last imaginary term vanishes because $V_+$ is orthogonal to $JV_+$.

What we have shown is sometimes expressed by the relation $U(n)\simeq SO(2n) \cap Sp(2n)$.  In \cite{Fujii:2013sra}
it is called {\em reality condition}.

\section{Evolution of the tangent space}

In the previous section we have discussed the vector space tangent to the thimble at the saddle point.  In order to
compute the vector space tangent at any other point of the thimble it is necessary to evolve a basis of vectors
according to Eq.~(18) of \cite{Cristoforetti:2012su}.  Such evolution preserves the orthogonality\footnote{Note
  that the evolved matrix $V_+(\tau)$ is {\em not} a basis of the eigenvectors of the Hessian $H(z(\tau))$,
  computed in the evolved configuration $z(\tau)\in\Gamma$.} of $V_+$ and $V_-:=JV_+$.  In fact, if we parametrize
with $\tau$ the curve of steepest descent that connects a generic point on the thimble with the saddle point at
$\tau\rightarrow\infty$, the evolution equation becomes:
\begin{equation}
\label{eq:ev}
V_+(\tau+\dt) = V_+(\tau) + \dt H(z(\tau)) V_+(\tau);
\end{equation}
the orthogonality of $V_+(\tau)$ and $V_-(\tau)$ is preserved at any $\tau$ because
\begin{eqnarray*}
V^T_+(\tau+\dt) J V_+(\tau+\dt) 
&=& 
V_+(\tau) J V_+(\tau) + \dt 
\left[ V^T_+(\tau) J H(z(\tau)) V_+(\tau) + V^T_+(\tau) H(z(\tau))^T J V_+(\tau)\right] = \\
&&= 
0 + \dt \left[ V^T_+(\tau) \{J,H\} V_+(\tau) \right] = 0.\\
\end{eqnarray*}
On the other hand, $V_+(\tau+\dt)$ and $V_-(\tau+\dt)$ are not orthonormal anymore. If we orthonormalize them
(e.g. with Gram-Schmidt, as it is done in \cite{Fujii:2013sra}), we obtain a new basis $V'_+(\tau+\dt)$, such that
$V_+=V'_+ W$, with $W$ $n\times n$ and upper triangular.  Then we can use the projector $P$, defined in
Eq.~(\ref{eq:P}), to define the matrix $U_+(\tau+\dt) := P V'_+(\tau+\dt)$.  Now $U_+(\tau)$ is unitary for all
$\tau$, in fact:
\begin{eqnarray}
P^{\dag} P &=& 1_{2n} + i J, \label{eq:PP} \\
(PV'_+)^{\dag} (PV'_+) &=& (V'_+)^T (P^{\dag} P) (V'_+) = (V'_+)^T (1_{2n} + i J) (V'_+) = \\
&=& (V'_+)^T \, 1_{2n} \, (V'_+)  +  i (V_+ W^{-1})^T \, J \, (V_+ W^{-1}) = 1_{2n} + 0.
\nonumber
\end{eqnarray}
In particular $\det(U_+) = e^{i\phi}$ and we have shown that the residual phase is actually a phase.

\section{Evolution by continuous orthogonalization}

Instead of evolving the vectors in $V_{+}(\tau)$ with Eq.~(\ref{eq:ev}), we can combine evolution and
orthonormalization at every step as prescribed by the Drury-Davey \cite{Drury1980133,Davey1983343} method of
continuous orthogonalization (see also \cite{Avitabile20101038} for a nice geometrical discussion and
generalization).  The evolution equation, with Euler method, is:
\begin{eqnarray}
\label{eq:co}
V_+(\tau+\dt) &=& V_+(\tau) + \dt (1-V_+(\tau) V_+(\tau)^T) H(z(\tau)) V_+(\tau) = \nonumber\\
&&= V_+(\tau) + \dt (V_-(\tau) V_-(\tau)^T) H(z(\tau)) V_+(\tau).
\end{eqnarray}
It is straightforward to check that, at all times $\tau$, both the vectors $V_+$ and $V_-$ remain orthonormal and
orthogonal to each other:
\begin{eqnarray}
V_+(\tau)^T V_+(\tau) &=& 1_n, \label{eq:V+V+} \\
V_-(\tau)^T  V_-(\tau) &=& (J V_+(\tau))^T  (J V_+(\tau)) = 1_n, \\
V_+(\tau)^T V_-(\tau) &=& V_+(\tau)^T J V_+(\tau) = 0. \label{eq:V+V-}
\end{eqnarray}
Now we can define a unitary matrix directly from $V_+(\tau)$:
\begin{equation}
\label{eq:U=PV}
U_+(\tau)=P V_+(\tau).
\end{equation}
Eq.~(\ref{eq:co}) implements an Iwasawa projection (equivalently, a Gram-Schmidt infinitesimal orthonormalization)
at every $\tau$, but it is much more expensive than Eq.~(\ref{eq:ev})\footnote{Note that the evolution defined by
  Eq.~(6) of \cite{Cristoforetti:2012uv} is not correct.  In order to ensure an orthogonal evolution one should use
  instead Eq.~(\ref{eq:co}) above.}.  In fact, the cost of implementing Eq.~(\ref{eq:co}) scales like $n^3$.
Eq.~(\ref{eq:co}) will be used, in the next section, to deduce a simple formula for the residual phase, but
eventually it will not be needed in the method that we propose.  We will use Eq.~(\ref{eq:co}) only to cross-check
the results obtained with our method.

\section{Computing the residual phase}

After the preparatory analysis of the previous sections, we come to the formula for the computation of the residual
phase, that is the main result of this paper.  We can assume to know the phase $\phi_0$ at the stationary point,
and we can also assume that this is attained for $\tau=\tau_{\infty}$ sufficiently large (i.e., $\det
U_+(\tau_{\infty})\simeq \lim_{\tau\rightarrow\infty} \det U_+(\tau) = e^{i\phi_0}$).  Therefore,
\begin{eqnarray*}
\log \det U_+(\tau) 
&\stackrel{Eq.(\ref{eq:U=PV})}{=}& 
\log \det \left [ P V_+(\tau) \right] =
\Tr \log \left [ P V_+(\tau) \right] =
\\ &=&
\int_{\tau_{\infty}}^{\tau} ds \, \Tr  \left [ (P V_+(s))^{-1} P \frac{d V_+(s)}{ds} \right]  
\qquad + \qquad i \phi_0 
\\ &\stackrel{U_+ \text{\tiny unitary}}{=}&
\int_{\tau_{\infty}}^{\tau} ds \, \Tr  \left [ (P V_+(s))^{\dag} P \frac{d V_+(s)}{ds} \right]
\qquad + \qquad i \phi_0
\\ &=&
\int_{\tau_{\infty}}^{\tau} ds \, \Tr  \left [ V_+(s)^T (P^{\dag} P) \frac{d V_+(s)}{ds} \right]
\qquad + \qquad i \phi_0
\\ &\stackrel{Eq.(\ref{eq:PP})}{=}&
\int_{\tau_{\infty}}^{\tau} ds \, \Tr  \left [ V_+(s)^T (1_{2n} + i J) \frac{d V_+(s)}{ds} \right]
\qquad + \qquad i \phi_0
\\ &\stackrel{Eq.(\ref{eq:co})}{=}&
\int_{\tau_{\infty}}^{\tau} ds \, \Tr  \left [ V_+(s)^T (1_{2n} + i J) 
\left ( V_-(s) V_-(s)^T H(s) V_+(s) \right) 
\right]  \qquad + \qquad i\phi_0
\\ &\stackrel{Eq.(\ref{eq:V+V-})}{=}&
i \int_{\tau_{\infty}}^{\tau} ds \, \Tr  \left [ V_+(s)^T \, J \, V_-(s) \, V_-(s)^T \, H(s) \, V_+(s) \right]  
\qquad + \qquad i\phi_0
\\ &\stackrel{Eq.(\ref{eq:V-})}{=}&
i \int_{\tau_{\infty}}^{\tau} ds \, \Tr  \left [ 
V_+(s)^T \, J^2 \, V_+(s) \, V_+(s)^T \, J^T \, H(s) \, V_+(s) 
\right]
\qquad + \qquad i\phi_0
\\ &\stackrel{J^2=-1, J^T=-J}{=}&
(-1)^2 i \int_{\tau_{\infty}}^{\tau} ds \, \Tr  \left [ V_+(s)^T \, V_+(s) \, V_+(s)^T \, J H(s) \, V_+(s) \right]
\qquad + \qquad i\phi_0
\\ &\stackrel{Eq.(\ref{eq:V+V+})}{=}&
i \int_{\tau_{\infty}}^{\tau} ds \, \Tr  \left [ V_+(s)^T \, J H(s) \, V_+(s) \right]  
\qquad + \qquad i\phi_0.
\end{eqnarray*}

Note that the result is purely imaginary, which confirms that the residual phase, in this setup, is indeed a
phase\footnote{Note also that the inverse matrix that appears in \cite{Cristoforetti:2012uv} has disappeared here,
  because the matrix $U_+(s)$ (called $T_{\phi_s}$ in \cite{Cristoforetti:2012uv}) is actually unitary.}.  As a
result, we have to compute the trace of the operator $JH(z)$ on the tangent space $T_z \Gamma$.

For very large $n$ it should be convenient to use a stochastic estimator of the trace, rather than compute it
fully.  Using $N_R$ random noises, we have:
\begin{eqnarray}
\label{eq:noise}
\Tr  \left [ V_+(s)^T \, J H(s) \, V_+(s) \right]  
&=&
\lim_{N_R\rightarrow\infty}
\frac{1}{N_R}\sum_{r=1}^{N_R} \xi^{(r) T} V_+(s)^T \, J H(s) \, V_+(s) \xi^{(r)}.
\end{eqnarray}
Note that the vectors $\eta^{(r)}(s) = V_+(s) \xi^{(r)}$ are generic random vectors in $T_{z(s)} \Gamma$.  One way
to compute Eq.~(\ref{eq:noise}) is by extracting random vectors $\eta^{(r)}(s) \in \mathbb{C}^n$, evolve them as
usual down to $\tau_{\infty}$ along the curve $z(\sigma)$, $\sigma\in[s,\tau_{\infty}]$, project them with the free
projector, evolve them back to $s$ and compute
\[
\frac{1}{N_R}\sum_{r=1}^{N_R} \eta^{(r) T}(s) \, J H(s) \, \eta^{(r)}(s).
\]
The evolution back and forth ensures the isotropy of the distribution of the $\eta$ \cite{Cristoforetti:2012su}.
But, note that we have to generate $\eta$ for each $s$.  The final formula is therefore:
\begin{equation}
\label{eq:final}
\log \det U_+(\tau) - i\phi_0 = 
\lim_{N_R\rightarrow\infty}
i \int_{\tau_{\infty}}^{\tau} ds \, 
\frac{1}{N_R}\sum_{r=1}^{N_R} \eta^{(r) T}(s) \, J H(s) \, \eta^{(r)}(s).
\end{equation}
The costs of computing Eq.~(\ref{eq:final}) scales as $n \times N_{\tau}^2 \times N_R$, where $N_{\tau}$ is the
number of steps in which the dimension along the gradient flow is subdivided.  This cost is a drastic improvement
over what we had estimated in \cite{Cristoforetti:2012su}.  Of course, one expects that the required $N_R$ will
increase linearly with $n$, but the experience with stochastic estimators tells that it is usually sufficient to
use $N_R \ll n$.

\section{Numerical tests}

In order to test our method, we implemented two algorithms.  One code computes the residual phase as defined in
Eq.~(\ref{eq:final}), in the previous section (hereafter called {\em stochastic} method).  Another code computes
the residual phase by evolving the basis $V_{+}(\tau)$ with Eq.~(\ref{eq:co}) and then computes the determinant
with the lapack function {\tt zgeevx} \cite{lapack} (hereafter called the {\em exact} method).  The exact method
has of course very limited applicability, as it scales as $O(n^3)$ (although it scales linearly with $N_{\tau}$)
and it is hardly parallelizable.  The exact method is used here only to test the stochastic method\footnote{The
  method employed in \cite{Fujii:2013sra} is similar to our exact method, but uses the evolution defined by
  Eq.~(\ref{eq:ev}), and integrated with the fourth-order Runge-Kutta scheme, rather than Eq.~(\ref{eq:co}).  The
  method of \cite{Fujii:2013sra} is probably the best compromise on small lattices, but, on large lattices, it is
  expected to scale less favorably than the stochastic method presented here.}.

\begin{figure}[ht]
\includegraphics{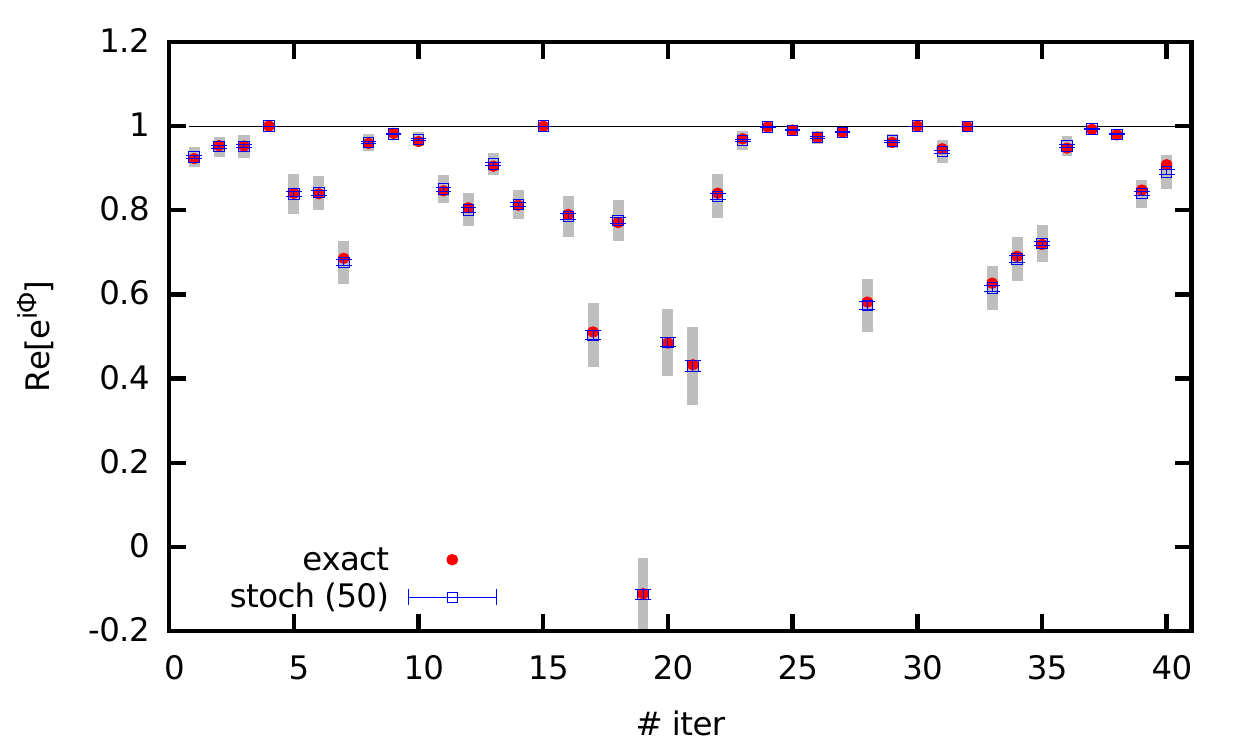}
\caption{\label{fig:correctness} Comparison of the computation of the residual phase with the {\em exact} and with
  the {\em stochastic} method.  Here we show the real part of the residual phase for a small sample of
  configurations that belong to the same Monte Carlo sequence (the iteration number is shown in the horizontal
  axis).  The blue errorbars are obtained by using $N_R=50$ stochastic estimators.  On the other hand, the grey
  band shows the standard deviation, which indicates the distribution that a single stochastic estimator would
  have.  The other parameters of the simulation are $n=2\times 16^2$, $N_{\tau}=64$, $\mu=1.05$, $m=\lambda=1$,
  $\Delta\tau=\Delta t_{Langevin}=10^{-3}$.  See \cite{Cristoforetti:2013wha} for the meaning of the parameters
  which are not defined here.  Note also that the configurations shown are not statistically significant:no attempt
  is made here to compute reliably the {\em average} residual phase for any ensemble.}
\end{figure}

In order to test the method, we have applied it to the usual complex scalar field with $\phi^4$ interaction and
with chemical potential \cite{Cristoforetti:2013wha}.  Here we limit ourself to a two dimensional system, in order
to sample very cheaply different lattices sizes.

As a test of correctness, in Fig.~\ref{fig:correctness} we compare the results of the two methods on a set of
configurations.  The test is passed brilliantly. Moreover, the grey bands in Fig.~\ref{fig:correctness} display the
standard deviation of the stochastic method; this is the statistical error that we expect if only one source is
used.  The size of the standard deviation indicates that even a single stochastic source is able to yield a fairly
accurate estimate in these cases.  As one can expect, the standard deviation is larger when the residual phase
differs more from its value at the saddle point, but the stochastic method seems always reliable.

\begin{figure}
\includegraphics[width=0.7\textwidth]{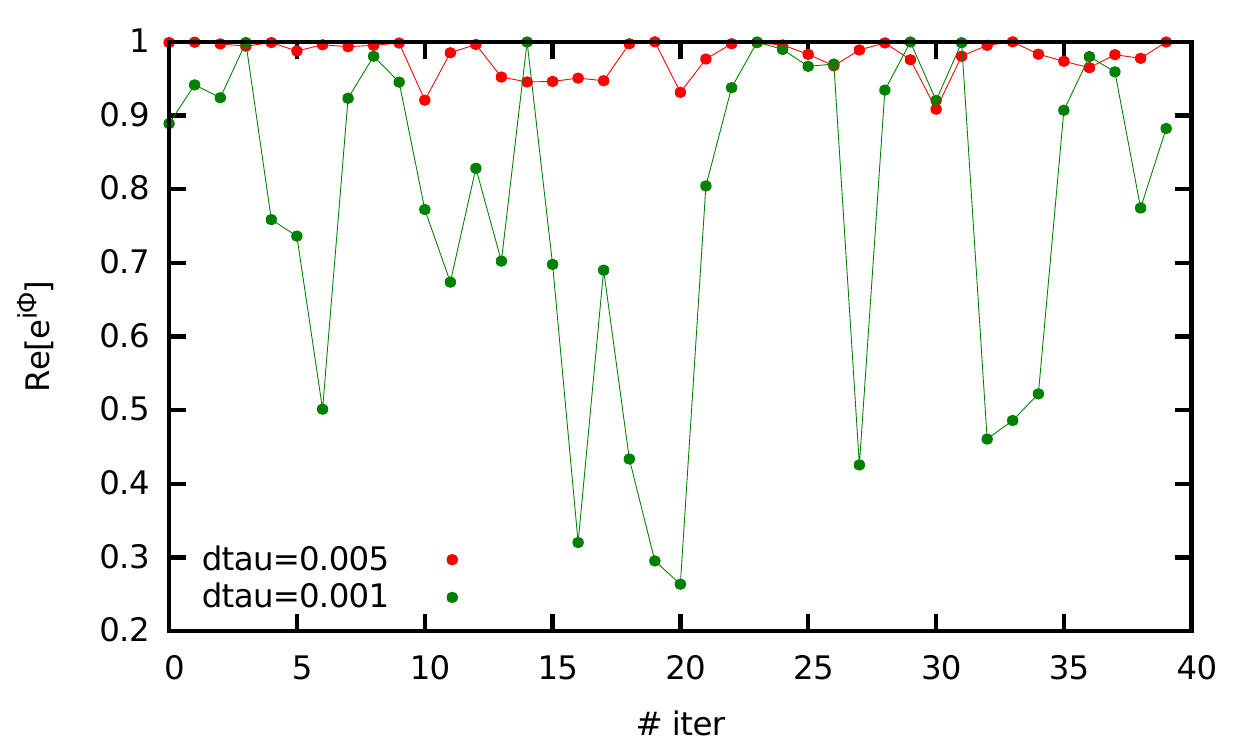}
\caption{\label{fig:totokyo} Residual phases for volume, $16^2$, $N_{\tau}=100$ and two different values of
  $d\tau$.  Already for $d\tau=5 \cdot 10^{-3}$, the residual phase come very close to those found in
  \cite{Fujii:2013sra}, which are obtained for $N_{\tau}=100$ and $d\tau=3 \cdot 10^{-2}$.}
\end{figure}

One might wonder whether the large deviations from $e^{i\phi} = 1$, in Fig.~\ref{fig:correctness}, are in conflict
with the very high average phases found in \cite{Fujii:2013sra}.  There is no conflict.  In fact, we recover the
agreement with the results in \cite{Fujii:2013sra}, if we use larger values of $\tau =N_{\tau} d\tau$.  This is
shown in Fig.~\ref{fig:totokyo}.  In particular, already for $N_{\tau}=100$ and $d\tau=5 \cdot 10^{-3}$, the phases
come very close to $e^{i\phi} = 1$.  However, the focus of this paper is on the precise determination of the
phases, and those largely deviating from unity are more interesting.

\begin{figure}[ht]
\includegraphics[width=0.7\textwidth]{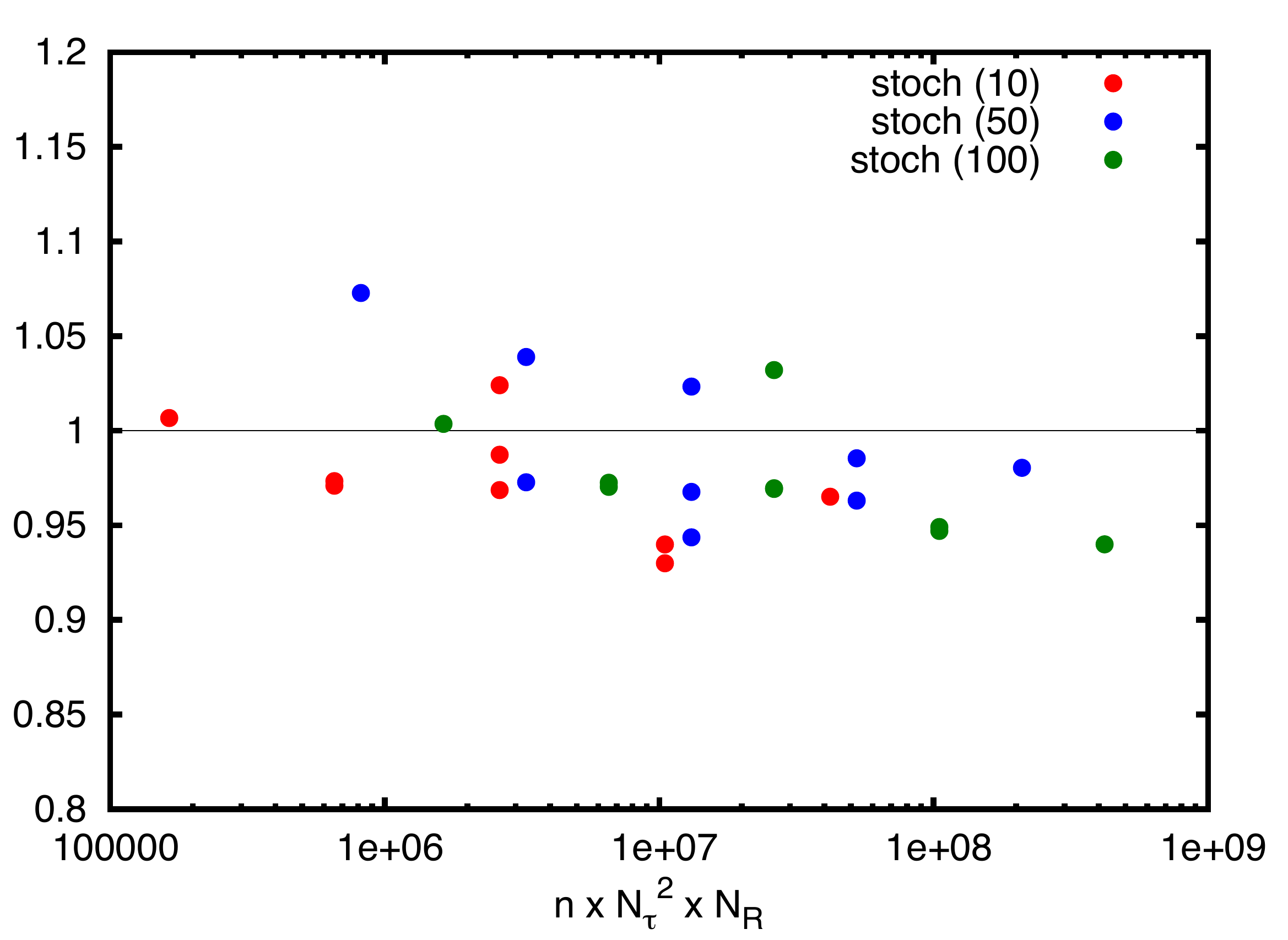}
\caption{\label{fig:costs} Scaling of the costs to compute the residual phase with the stochastic method.  The plot
  compares the actual cost with the estimate based on the scaling $O(n \times N_{\tau}^2 \times N_R)$, and
  normalized at the point with $n=2\times 4^2$, $N_R=1$, $N_{\tau}=32$.  The other parameters are the same as in
  the previous plot.  Different colors represent different $N_R$, which are given in the parenthesis on the
  top-right corner.}
\end{figure}

Besides these tests of correctness, we also tested the expected scaling of the computational costs (although still
on small lattices).  Fig.~\ref{fig:costs} shows that, as expected, the costs of the stochastic method scale as $O(n
\times N_{\tau}^2 \times N_R)$.  In these very small lattices, the exact and stochastic methods still have roughly
comaprable costs: for example, for $N_{\tau}=128$ and $n=2\times 16^2$, the exact method costs as much as the
stochastic one with $N_R\simeq 80$.  However, the stochastic method will necessarily become more efficient on
larger systems.  It is difficult to tell how the number of stochastic estimators $N_R$ will scale on large systems,
when keeping the precision of the computation of the residual phase fixed.  Generically, one expects a volume
dependence also in $N_R$, but the fact that $N_R=1$ seems already sufficient here is very encouraging.

In this paper, we did not try to estimate the {\em average} residual phase for any ensemble.  This paper is only
concerned with the proposal and the testing of a new method to compute the residual phase efficiently and precisely
on a given configuration.  In particular, the configurations in Fig.~\ref{fig:correctness} are not typical of their
ensemble.  Of course, it will be eventually very interesting to see how the average residual phase scales on larger
volumes and for realistic systems, and how it depends on the technical parameters that describe the thimble.  These
will be the crucial questions when studying a specific physical system, but they go beyond the scope of the present
paper.

\section{Conclusions}

In this paper we have proposed a new method to compute the residual phase that appears on Monte Carlo calculations
on a Lefschetz thimble.  In particular, our main result is the derivation of the formula in Eq.~(\ref{eq:final}).
By this, we have also corrected a mistake in \cite{Cristoforetti:2012uv}.  Moreover, we have reported the results
of the testing of our code and we have also verified the expected scaling of the costs as $O(n \times N_{\tau}^2
\times N_R)$.  A further advantage of the stochastic method is the fact that it can be easily parallelized across
the physical volume (our code is indeed already parallel).  The fact that already one single stochastic estimator
seems to be sufficient in these tiny lattices is certainly not conclusive, but quite encouraging.

\begin{acknowledgments}
It is a pleasure to thank Y.~Kikukawa and H.~Fujii for very interesting discussions.  This research is supported in
part by the INFN SUMA project, by the Research Executive Agency (REA) of the European Union under Grant Agreement
No.  PITN-GA-2009-238353 (ITN STRONGnet), and by the INFN i.s. MI11 and by MIUR contract PRIN2009
(20093BMNPR\_004).
\end{acknowledgments}

\appendix
\section{The residual determinant in the Metropolis algorithm}
\label{sec:apx}

In this section, we extend our analysis to the algorithm discussed in \cite{Mukherjee:2013aga}.  In that case, the
manifold $\Gamma$ is explored by making proposals that are uniform in the variables $\eta \in \mathbb{R}^n$, that
diagonalize and rescale the quadratic part of the action. Therefore, the residual Jacobi determinant in that case
is:
\begin{equation}
\det \left ({\mathbf{J}^{\phi}_{\eta}} \right ) = \det \left ({\frac{\partial \phi}{\partial \eta}} \right )
\label{eq:metdet}
\end{equation}
which is not a pure phase, in general, but rather a {\em residual determinant}.  However, we show in the following
that the {\em phase} of the determinant in Eq.~(\ref{eq:metdet}) is the same as the phase of $\det (U_+)$
discussed, for the Langevin algorithm, in the main text of this paper.

Indeed, the evolution equation for $\mathbf{J}^{\phi}_{\eta}$ is
\begin{equation}
\frac{d\mathbf{J}^{\phi}_{\eta}}{d \tau} = 
\overline{\partial^2_{\mathbf{\phi}^2} S} \overline{\mathbf{J}^{\phi}_{\eta}},
\nonumber
\end{equation}
with the boundary conditions
\begin{equation}
\left [\mathbf{J}^{\phi}_{\eta} \right ]_{ij} (\tau \to -\infty) = u^{(i)}_j,
\nonumber
\end{equation}
where the $u^{(i)}_j$ are the same of Sec.~\ref{sec:saddle}.  Now, from Eq.~(\ref{eq:ev}) we get
\begin{equation}
\frac{d PV_+}{d \tau} 
= PHV_+ 
= \overline{\partial^2_{\mathbf{\phi}^2} S} \overline{PV_+},
\nonumber
\end{equation}
where $\overline{P}$ is the $n\times 2n$ matrix $(1_n \, , \; \; -i 1_n)$, and we have exploited the identity
\begin{equation}
PH = 
\left (\partial^2_{\phi_R^2} S_R+i \partial^2_{\phi_R \phi_I} S_R \, , \quad  
\partial^2_{\phi_R \phi_I} S_R -i \partial^2_{\phi_R^2} S_R \right)
= \overline{\partial^2_{\mathbf{\phi}^2} S} \overline{P},
\nonumber
\end{equation}
where each block in the central term is an $n \times n$ complex matrix.  Thus, $PV_+$ and
$\mathbf{J}^{\phi}_{\eta}$ have identical evolution equations.  Their boundary conditions are also identical, as
evident from Eq.~(\ref{eq:P}).  Thus, the matrix $\mathbf{J}^{\phi}_{\eta}$ defined in \cite{Mukherjee:2013aga} is
identical to $PV_+$.

If the aforesaid mapping between the $\eta$ and the $\phi$ variables exists, then $\mathbf{J}^{\phi}_{\eta}$
is invertible. In which case it can be uniquely decomposed as
\begin{equation}
\mathbf{J}^{\phi}_{\eta} = U_+ W,
\nonumber
\end{equation}
where $U_+$ is unitary and $W$ is upper-triangular with \emph{real} diagonal elements.  Therefore the phase of the
residual determinant $\det (\mathbf{J}^{\phi}_{\eta})$ is simply $\arg (\log \det U_+)$, which is exactly the
residual phase that we get for the algorithm described in \cite{Mukherjee:2013aga}.

\bibliography{../density}{}
\end{document}